# O(lgN) Line Clipping Algorithm in $E^2$

Václav Skala[1]
Department of Informatics and Computer Science
University of West Bohemia
Univerzitní 22, Box 314, 306 14 Plzen
Czech Republic
e-mail: skala@kiv.zcu.cz    http://herakles.zcu.cz/~skala

**Abstract**

A new $O(\lg N)$ line clipping algorithm in $E^2$ against a convex window is presented. The main advantage of the presented algorithm is the principal acceleration of the line clipping problem solution. A comparison of the proposed algorithm with others shows a significant improvement in run-time. Experimental results for selected known algorithms are also shown.

## 1. Introduction

Many algorithms for clipping lines against convex or non-convex windows in $E^2$ with many modifications derived from well known Cohen-Sutherland's, Liang-Barsky's [LIA83a],[LIA84a] and Cyrus-Beck's [CYR79a] algorithms have been published. All of them have the same complexity $O(N)$, with an exception of Rappaport's algorithm [RAP91a] which has $O(\lg N)$ complexity. Their speed is determined by more or less clever implementation of tests and intersection computation. The convexity feature of the clipping polygon and the possibility of binary search usage over polygon vertices, because of known vertices order, have been used for principal speed up of the ECB line clipping algorithm [SKA93b] that resulted into new line clipping algorithm with complexity $O(\lg N)$. It has been expected that an algorithm for line clipping against convex polygon with complexity $O(\lg N)$ exists, see [CHA87a]. An algorithm for a line segment clipping with $O(\lg N)$ complexity was published in [RAP91a]. The known algorithms for clipping lines against a general convex window do not make tests similar to Cohen-Sutherland's clipping algorithm. The main reason seems to be the computational cost of such tests for convex windows. If a clipping algorithm is to be effective, it is necessary to distinguish cases where lines pass through a given window from those where lines do not intersect the window. Cyrus-Beck's (CB) algorithm solves this problem by direct computation of points of intersections, the ECB algorithm uses the separation theorem for Cyrus-Beck's algorithm to achieve a speed up of approx. 1.2 - 2.5 times. Cyrus-Beck's (CB), Efficient Cyrus-Beck's (ECB) and Rappaport's algorithms have been compared with the new proposed $O(\lg N)$ algorithm.

The ECB algorithm does not use the known order of vertices of the given clipping polygon for a principal speed up of the algorithm, though it has the complexity $O(N)$.

The Rappaport's algorithm [RAP91a] is the only one algorithm with $O(\lg N)$ complexity that could be used for line segments clipping against convex polygon. The algorithm, see alg.1, is based on known fact that an answer whether a point is inside of the convex polygon can be given in $O(\lg N)$ steps, where N is a number of vertices of the given polygon [PRE85a].

---

[1] Supported by the grant UWB-156/1995
Published in Computers & Graphics, Pergamon, No.4, Vol.18, pp.517-524, 1994.

**procedure** RAPPAPORT ($\mathbf{x}_A$, $\mathbf{x}_B$);
{ $\mathbf{x}_A$, $\mathbf{x}_B$ are end-points of the clipped line segment }
**begin  if** CLASSIFY ($\mathbf{x}_A$) = IN **then**
    **begin** (s,s1) := SECTOR ($\mathbf{x}_A$, $\mathbf{x}_B$);
        **if** $\mathbf{x}_B$ is to the left of s-s1 edge of the polygon { s1 is the next vertex to vertex s }
        **then**   OUTPUT ($\mathbf{x}_B$) { the line segment is totally inside }
        **else**
        **begin**  compute the intersection point of the line segment with the edge s-s1 ($\mathbf{x}$);
            OUTPUT($\mathbf{x}$);
        **end**
    **end**
    **else**
    **begin** (left_sup,right_sup) := SUPPORT_VERTICES ($\mathbf{x}_A$);
        **if** $\mathbf{x}_B$ is left of left_sup or right of right_sup
        **then**   DO_NOTHING
        **else**   **begin**  { find an intersected edge from the front chain }
            (s,s1) := FRONT_SECTOR (left_sup,right_sup);
            **if** $\mathbf{x}_B$ is to the right of s-s1
            **then** DO-NOTHING
            **else**
            **begin**  compute the intersection point of the line segment with
                the edge s-s1 ($\mathbf{x}$);
            OUTPUT ($\mathbf{x}$);
            (s,s1) := BACK_SECTOR (right_sup,left_sup);
            **if** $\mathbf{x}_B$ is to the left of s-s1
            **then** OUTPUT ($\mathbf{x}_B$)
            **else**
            **begin**  { find an intersected edge from the back chain }
                compute the intersection point of the line segment
                  with the edge s-s1 ($\mathbf{x}$);
                OUTPUT ($\mathbf{x}$);
            **end**
            **end**
        **end**
    **end**
**end** { RAPPAPORT };

Algorithm 1

There are used the following functions in alg.1:
- CLASSIFY ($\mathbf{x}$) gives an answer if the point $\mathbf{x}$ is inside of the given convex polygon in $O(\lg N)$ steps and has complexity { ( := , < , ± , * , / ) counting FPP operations only }

$$(0,2,4,4,0) + (0,1,2,2,0) * \lg N$$

- SECTOR ($\mathbf{x}_A$, $\mathbf{x}_B$); finds an edge with vertices (s, s1) that is intersected by the given line segment $\mathbf{x}_A$, $\mathbf{x}_B$ in $O(\lg N)$ steps and has complexity

$$(7,2,9,5,0) * \lg N$$

- SUPPORT_VERTICES ( $\mathbf{x}_A$ ) finds the (left_sup, right_sup) indexes of end-points of the back and front chains that are formed by edges of the given polygon in $O(\lg N)$ steps and has complexity

$$(0,2,10,4,0) + (0,2,10,4,0)*\lg N$$

- FRONT_SECTOR (left_sup, right_sup) finds from front chain of edges with vertices (s, s1) that is intersected by the given line segment $\mathbf{x}_A$, $\mathbf{x}_B$ in $O(\lg N)$ steps and has complexity

$$(0,1,2,2,0)*\lg N$$

- BACK_SECTOR (left_sup, right_sup) finds from back chain of edges with vertices (s, s1) that is intersected by the given line segment $\mathbf{x}_A$, $\mathbf{x}_B$ in $O(\lg N)$ steps

$$(0,1,2,2,0)*\lg N$$

It can be seen that all steps are of $O(\lg N)$ complexity and therefore the whole algorithm is of $O(\lg N)$ complexity, too. Unfortunately some steps are quite complex and the overall complexity for the worst case can be estimated as

$$(4,2,12,22,2) + (0,4,14,8,0)*\lg N$$

Detailed description of the Rappaport's algorithm can be found in [RAP91a].

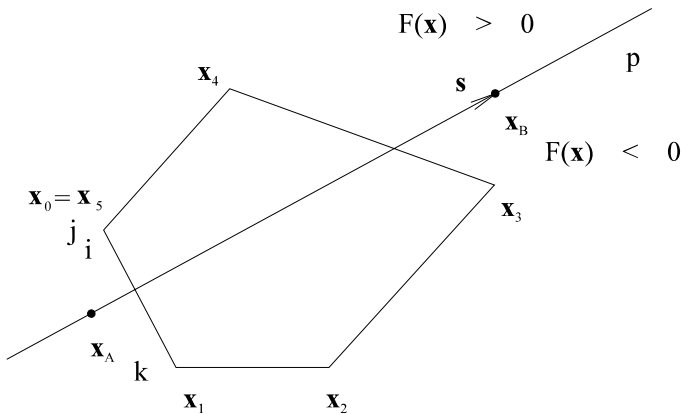
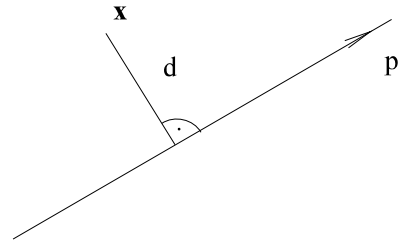

Figure 1    Figure 2

## 2. Proposed algorithm

Let us suppose that we have a given convex clipping polygon anti-clockwise oriented and line $p$ is determined by two end-points

$$\mathbf{x}_A = [x_A, y_A]^T \quad , \quad \mathbf{x}_B = [x_B, y_B]^T$$

The convex window is represented by $n+1$ points

$$\mathbf{x}_i = [x_i, y_i]^T, \quad i = 0, ..., n$$

where: points $\mathbf{x}_0$ and $\mathbf{x}_n$ are identical (column notation is used), $x_i$ and $y_i$ are coordinates of the vertex $\mathbf{x}_i$.

The notation $\overline{\mathbf{x}_i \mathbf{x}_k}$ is used for a polyline from $\mathbf{x}_i$ to $\mathbf{x}_k$, i.e. it is a chain of line segments from $\mathbf{x}_i$ to $\mathbf{x}_k$.

Let us define the separation function $F(\mathbf{x})$ in the form

$$F(\mathbf{x}) = Ax + By + C$$

where $F(\mathbf{x}) = 0$ is an equation for the given line $p$ and assume that the line has the orientation shown in Fig.1, $\mathbf{x}$ is defined as $\mathbf{x} = [x, y]^T$.

It can be seen, Fig.2, that the oriented distance $d$ of the point $\mathbf{x}$ from the line $p$ can be determined as

$$d = \frac{Ax + By + c}{\sqrt{A^2 + B^2}}$$

It means that the value of the function $F(\mathbf{x})$ is actually proportional to the distance $d$ for the given line $p$. First of all, let us assume that (see Fig.1)

$$i = 0 \ ; \qquad j = n \ ; \qquad k = (i + j) \ \mathbf{div} \ 2 \ ;$$

and

$$\mathbf{x}_0 \equiv \mathbf{x}_n \qquad \mathbf{x}_i = \mathbf{x}_0 \qquad \mathbf{x}_j = \mathbf{x}_n \qquad \mathbf{x}_k \equiv \mathbf{x}_2$$

Let us concentrate on a special case shown in Fig.1. If the points $\mathbf{x}_i$ and $\mathbf{x}_k$ are on the opposite sides of the line $p$, i.e.

$$F(\mathbf{x}_i) * F(\mathbf{x}_k) < 0$$

then there must be just one intersection point on the chains $\overline{\mathbf{x}_i \mathbf{x}_k}$ and $\overline{\mathbf{x}_k \mathbf{x}_j}$ for each chain, because the given polygon is convex. Because $F(\mathbf{x}_i) * F(\mathbf{x}_k) < 0$ for the chain $\overline{\mathbf{x}_i \mathbf{x}_k}$ there must exist an index $l$ so that

$$F(\mathbf{x}_l) * F(\mathbf{x}_{l+1}) < 0 \qquad i \leq l < k$$

i.e. an edge $\overline{\mathbf{x}_l \mathbf{x}_{l+1}}$ must be intersected.

Similarly for the chain $\overline{\mathbf{x}_k \mathbf{x}_j}$. It is obvious that in this case the intersection point can be found in $O(\lg M)$ steps using binary search over vertices, where M is a number of line segments in the given chain.

Unfortunately, other possible situations are more complex to solve, see Fig.3. It is possible to distinguish four fundamental cases supposing the previously shown orientation of the separation function $F(\mathbf{x})$. In case a) the chain $\overline{\mathbf{x}_k \mathbf{x}_j}$ can be removed, while in case b) the chain $\overline{\mathbf{x}_i \mathbf{x}_k}$ can be removed. In the first, resp. second, case index j, resp. index i, must be changed to k. In both cases a new value of k must be computed as

$$k = (i + j) \ \mathbf{div} \ 2$$

Both mentioned cases can be distinguished by a criterion

$$F(\mathbf{x}_{i+1}) < F(\mathbf{x}_i)$$

because if $F(\mathbf{x}_{i+1}) < F(\mathbf{x}_i)$ then the chain $\overline{\mathbf{x}_i \mathbf{x}_k}$ can intersect the line $p$, see Fig.3. This condition actually expresses that we are getting closer to the line $p$, i.e. the oriented distance $d$ is smaller.

In both cases we assumed that the line p has the shown orientation, i.e. $F(\mathbf{x}_i) > 0$ and

$$F(\mathbf{x}_i) \leq F(\mathbf{x}_k)$$

Possible situations as a variation of cases a) and b) in Fig.3, when this condition is not true, are shown as cases c) and d).

A little bit more complex situation is shown by cases c) and d) where $F(\mathbf{x}_i) > F(\mathbf{x}_k)$. In case c) the chain $\overline{\mathbf{x}_k\mathbf{x}_j}$ can be removed, while in case d) the chain $\overline{\mathbf{x}_i\mathbf{x}_k}$ can be removed. In the first, resp. second, case index j, resp. index i, must be changed to k. In both cases a new value of k must be again determined as

$$k = (i + j) \text{ div } 2$$

Both last mentioned cases can be distinguished by using criterion

$$F(\mathbf{x}_{k+1}) < F(\mathbf{x}_k)$$

Actually we must distinguish whether we are getting closer to the given line $p$ or not. If the line $p$ has an opposite orientation then similar situations must be solved, see Algorithm 2.

This procedure is repeated until

$$F(\mathbf{x}_i) * F(\mathbf{x}_k) < 0$$

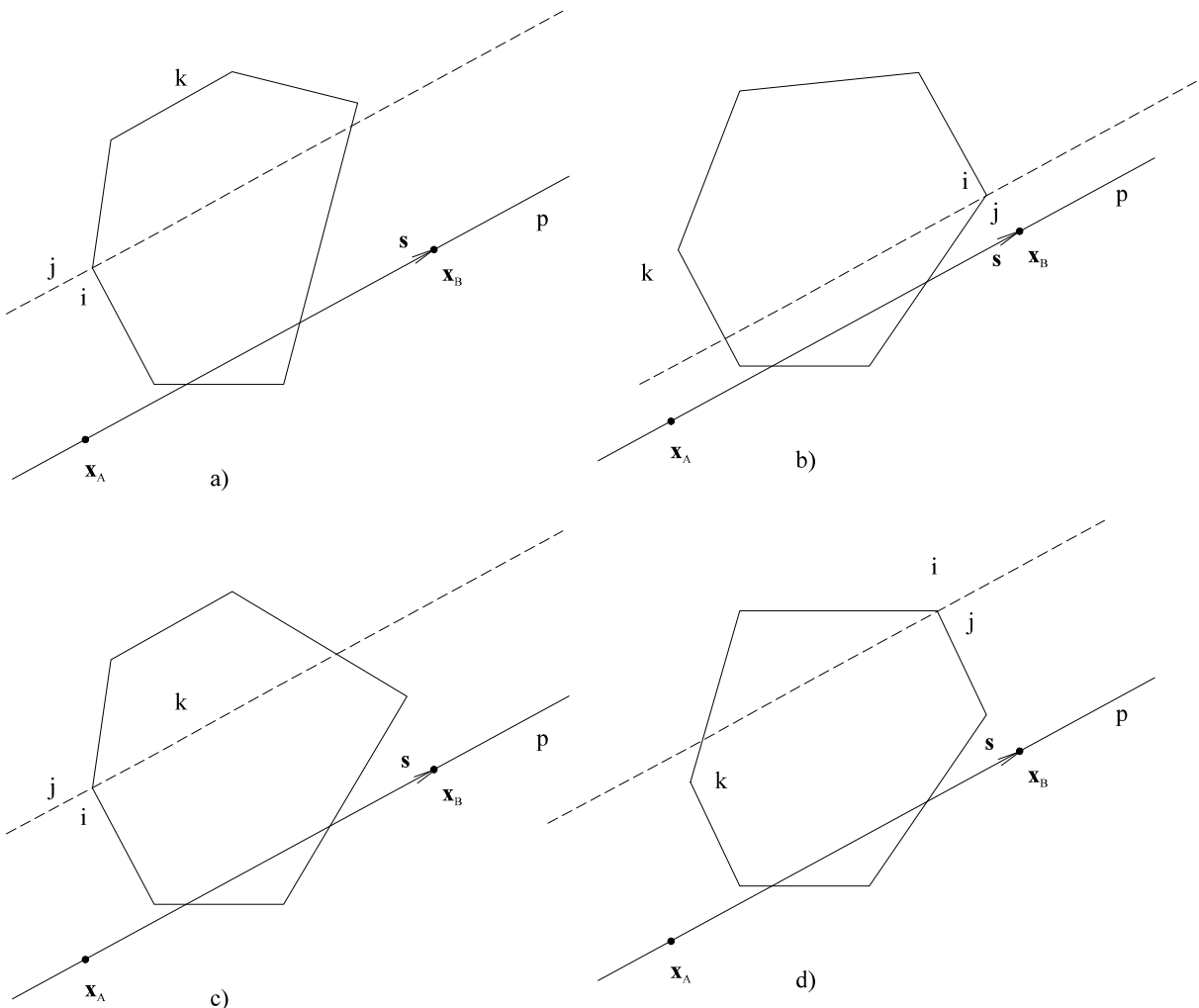

Dashed lines mean points $\mathbf{x}$, where $F(\mathbf{x}) = F(\mathbf{x}_i)$

Figure 3

If this condition becomes true we will obtain two chains $\overline{\mathbf{x}_i \mathbf{x}_k}$ and $\overline{\mathbf{x}_k \mathbf{x}_j}$ that intersect the line $p$ and binary search over vertices can be used again as we get a similar situation shown in Fig.1.

Now it can be seen that all parts of the proposed algorithm are of complexity $O(\lg M)$, where M is a number of edges in the given chain because we used for all steps the binary search over vertices of the clipping convex polygon. The whole proposed $O(\lg N)$ algorithm is described by Algorithm 2. It is necessary to point out that for effective implementation values $F(\mathbf{x}_i)$ should be stored in separate variables as they are used several times.

**procedure** CLIP 2D lg ( $\mathbf{x}_A$, $\mathbf{x}_B$ );
{ Note: initialization of the clipping window $\mathbf{x}_n := \mathbf{x}_0$ }
**function macro** F( **x** ): **real**;
{ should be implemented as an in-line function }
**begin**
    F := A * x + B * y + C;
**end** { F };

**function** SOLVE ( i , j ): **real**;
{ finds two nearest vertices on the opposite sides }
{ of the given line $p$ }
**begin**   **while** ( j - i ) $\geq$ 2 **do** { j $\geq$ i always }
    **begin**   k := ( i + j ) **div** 2; { shift to the right }
        **if** $(F(\mathbf{x}_i) * F(\mathbf{x}_k)) < 0$ **then** j := k **else** i := k;
    **end** { while };
    SOLVE := INTERSECTION ( $p$ , $\mathbf{x}_i$ , $\mathbf{x}_j$ );
    { gives the value t of an intersection point }
    { of the line p with the given line segment $\mathbf{x}_i \mathbf{x}_j$ }
**end** { SOLVE };

**begin** { determine the A, B, C values for the function $F(\mathbf{x})$ }
    $A := y_1 - y_2$;     $B := x_2 - x_1$;     $C := x_1 * y_2 - x_2 * y_1$;
    i := 0;         j := n;
    { for lines             $t_{min} := -\infty$;     $t_{max} := \infty$; }
    { for line segments   $t_{min} := 0$;     $t_{max} := 1$; }
    **while** ( j - i ) $\geq$ 2 **do**
    **begin**   k := ( i + j ) **div** 2; { shift to the right }
        **if** $(F(\mathbf{x}_i) * F(\mathbf{x}_k)) < 0$ **then**
        **begin** { see fig.1 }
            $t_1$ := SOLVE ( i , k ) ; { find an intersection on $\overline{\mathbf{x}_i \mathbf{x}_k}$ chain }
            $t_2$ := SOLVE ( k , j ); { find an intersection on $\overline{\mathbf{x}_k \mathbf{x}_j}$ chain }
            { for the line segment clipping include the next 5 lines         }
            { **if** $t_1 > t_2$ **then begin** t := $t_2$ ; $t_2$ := $t_1$ ; $t_1$ := t **end;**        }
            {compute $<t_1, t_2>$   as   $<t_1, t_2> \cap <0, 1>$           }
            { $t_1$ := max ($t_{min}$ , $t_1$ );    $t_2$ := min ($t_{max}$ , $t_2$ );        }
            { **if** $<t_1, t_2> \geq \varnothing$ **then** draw line segment          }
            { **if** $t_1 \leq t_2$ **then**   SHOW-LINE( $\mathbf{x}(t_1)$ , $\mathbf{x}(t_2)$ );         }

```
                    EXIT { exit procedure CLIP 2D lg };
            end { if };

            { for the polygon orientation shown in fig.3 }
            if F(x_i) > 0  then
            begin  { for the orientation of line  p  shown in fig.3 }
                    if  F(x_i) < F(x_k)  then { cases a and b }
                    begin { DELETE CHAIN ( i , j ) removes the chain x x }
                            if  F(x_{i+1}) < F(x_i)  then
                                    begin j := k; { DELETE CHAIN ( k , j ); case a } end else
                                    begin i := k; { DELETE CHAIN ( i , k ); case b } end
                    end
                    else { cases c and d }
                    begin
                            if  F(x_{k+1}) > F(x_k)  then
                            begin j := k; { DELETE CHAIN ( k , j ); case c } end else
                            begin i := k; { DELETE CHAIN ( i , k ); case d } end
                    end
            end

            else
            begin  { for an opposite orientation of the line   }
                    if  F(x_i) > F(x_k)  then
                    begin  if  F(x_{i+1}) > F(x_k)  then
                                    begin  j := k; { DELETE CHAIN ( k , j ); } end
                           else
                                    begin i := k; { DELETE CHAIN ( i , k ); } end
                    end
                    else
                    begin  if  (F(x_{k+1}) * F(x_k)) < 0  then
                                    begin j := k; { DELETE CHAIN ( k , j ); } end
                           else
                                    begin i := k; { DELETE CHAIN ( i , k ); } end
                    end
            end
        end { while }
end { CLIP-2D-lg }
```

Algorithm 2

## 3. Theoretical analysis and experimental results

Before making any experiments it is convenient to point out that time needed for operations ( $:=$ , $<$ , $\pm$ , $*$ , $/$ ) differ significantly from computer to computer.

| float | $:=$ | $<$ | $\pm$ | $*$ | $/$ |
|---|---|---|---|---|---|
| time | 33 | 50 | 16 | 20 | 114 |

Table 1

Let us introduce coefficients of the effectivity $v$ as

$$v_1 = \frac{T_{CB}}{T} \qquad v_2 = \frac{T_{CB}}{T_0} \qquad v_3 = \frac{T_R}{T}$$

where: $T_{CB}$, $T_0$, $T_R$, $T$ are execution times needed by Cyrus-Beck's (**CB**), **ECB**, Rappaport's and proposed $O(\lg N)$ algorithms.

Description of **CB** and **ECB** algorithms can be found in [SKA93b] together with their theoretical and experimental comparisons.

Generally it is possible to express the complexity of the **CB** algorithm

$$(8,3,6,4,0) + (5,3,7,4,1) * N$$

and time of computation as $T_{CB}$ (for PC 486, see tab.1) can be estimated

$$T_{CB} = 590 + 621 * N$$

The complexity of the **ECB** algorithm (in the worst case) as

$$(15,3,11,14,2) + (3,1,1,3,0) * N$$

and time of computation $T_0$ can be estimated as

$$T_0 = 1329 + 257 * N$$

Description of **CB** and **ECB** algorithms and their theoretical and experimental comparisons can be found in [SKA93b]. Their complexities are $O(N)$.

Complexity of the **Rappaport's algorithm** can be expressed as

$$(4,2,12,22,2) + (0,4,14,8,0) * \lfloor \lg(N+1) \rfloor$$

and time of computation $T_k$ can be estimated as

$$1092 + 584 * \lfloor \lg(N+1) \rfloor$$

while for the suggested algorithm $O(\lg N)$ the complexity is given as

$$(14,4,11,15,2) + (2,4,6,6,0) * \lfloor \lg(N+1) \rfloor$$

and time of computation $T$ can be estimated as

$$T = 1267 + 376 * \lfloor \lg(N+1) \rfloor$$

The Rappaport's and proposed algorithms are of $O(\lg N)$ complexity. Theoretical speed up is given in tab.2 (the worst cases and operations in floating point were considered only)

| N | 4 | 5 | 6 | 7 | 8 | 9 | 10 | 20 | 30 | 50 | 100 |
|---|---|---|---|---|---|---|---|---|---|---|---|
| $v_1$ | 1.28 | 1.54 | 1.80 | 2.06 | 2.01 | 2.23 | 2.45 | 4.13 | 6.11 | 8.98 | 16.08 |
| $v_2$ | 0.98 | 1.09 | 1.20 | 1.31 | 1.22 | 1.31 | 1.41 | 2.06 | 2.87 | 4.02 | 6.93 |
| $v_3$ | 1.19 | 1.19 | 1.19 | 1.19 | 1.24 | 1.24 | 1.24 | 1.27 | 1.27 | 1.30 | 1.33 |

Theoretical estimations (worst case)
Table 2

| $v_1$ | 3 | 4 | 5 | 6 | 7 | 8 | 9 | 10 | 30 | 50 | 100 |
|---|---|---|---|---|---|---|---|---|---|---|---|
| 0% | 1.00 | 1.48 | 1.26 | 1.38 | 1.47 | 1.68 | 1.86 | 1.52 | 3.70 | 6.28 | 10.42 |
| 20% | 0.93 | 0.91 | 1.05 | 1.24 | 1.30 | 1.33 | 1.99 | 2.07 | 4.47 | 6.11 | 9.28 |
| 40% | 1.01 | 1.23 | 1.11 | 1.36 | 1.34 | 1.48 | 1.19 | 2.30 | 3.53 | 6.06 | 10.20 |
| 60% | 1.09 | 1.19 | 1.35 | 1.32 | 1.30 | 1.58 | 1.42 | 1.57 | 3.44 | 6.10 | 10.18 |
| 80% | 0.82 | 1.23 | 1.06 | 1.14 | 1.46 | 1.45 | 1.64 | 2.14 | 3.74 | 6.03 | 10.85 |
| 100% | 0.80 | 1.02 | 1.08 | 1.11 | 1.40 | 1.61 | 1.23 | 1.61 | 4.40 | 5.80 | 11.11 |

| $v_2$ | 3 | 4 | 5 | 6 | 7 | 8 | 9 | 10 | 30 | 50 | 100 |
|---|---|---|---|---|---|---|---|---|---|---|---|
| 0% | 1.47 | 1.81 | 1.81 | 1.89 | 1.12 | 1.77 | 1.89 | 1.61 | 2.00 | 2.29 | 2.37 |
| 20% | 1.19 | 1.27 | 1.40 | 1.81 | 1.66 | 1.61 | 1.70 | 3.28 | 1.96 | 1.96 | 1.98 |
| 40% | 1.33 | 1.27 | 1.19 | 1.39 | 1.37 | 1.72 | 1.79 | 1.90 | 1.95 | 1.91 | 2.03 |
| 60% | 1.17 | 1.14 | 1.33 | 1.52 | 1.38 | 1.73 | 1.55 | 1.47 | 1.70 | 2.06 | 2.13 |
| 80% | 0.91 | 1.22 | 1.40 | 1.24 | 1.62 | 1.79 | 1.32 | 1.63 | 1.86 | 2.12 | 2.21 |
| 100% | 0.98 | 1.14 | 1.35 | 1.26 | 1.49 | 1.75 | 1.54 | 1.46 | 2.14 | 2.14 | 2.35 |

| $v_3$ | 3 | 4 | 5 | 6 | 7 | 8 | 9 | 10 | 30 | 50 | 100 |
|---|---|---|---|---|---|---|---|---|---|---|---|
| 0% | 2.96 | 3.44 | 2.90 | 2.62 | 2.68 | 2.83 | 2.78 | 1.91 | 2.22 | 2.44 | 2.13 |
| 20% | 3.76 | 1.98 | 2.24 | 2.26 | 2.41 | 2.01 | 2.85 | 2.96 | 2.64 | 2.52 | 2.15 |
| 40% | 2.82 | 2.65 | 2.56 | 2.89 | 2.36 | 2.25 | 1.71 | 3.20 | 2.42 | 2.43 | 2.26 |
| 60% | 3.13 | 2.81 | 2.67 | 2.59 | 2.50 | 2.44 | 2.07 | 2.30 | 2.34 | 2.31 | 2.07 |
| 80% | 2.70 | 3.10 | 2.29 | 2.27 | 2.35 | 1.97 | 2.46 | 2.49 | 2.30 | 2.09 | 2.39 |
| 100% | 2.40 | 2.53 | 2.12 | 1.96 | 2.25 | 2.43 | 1.68 | 2.24 | 2.19 | 1.99 | 2.12 |

Table 3

The proposed algorithm has been tested against Cyrus-Beck's, **ECB** and Rappaport's algorithms on data sets of line segments ($10^3$) with end points that have been randomly and uniformly generated inside a circle in order to eliminate an influence of rotation. Convex polygons were generated as N-sided convex polygons inscribed into a smaller circle.

There are practically no significant differences as far as the percentage is intersecting lines is concerned, see tab.3.

It can be seen that, see tab.3, that the proposed algorithm is significantly faster then CB algorithm. A comparison of **ECB** and proposed algorithms shows that for $N < 7$ the **ECB** algorithm is faster than the proposed one. "Waves" for $v_2$ are caused by the influence of binary division of an index interval and relation between data and convex polygon position. The waves can be seen in tab.2 with theoretical estimations, too. The significant difference for $N = 100$ is caused by considering the worst cases only in theoretical estimations.

The proposed $O(\lg N)$ algorithm is approx. two times faster than Rappaport's algorithm and it is much more simple to implement.

It is necessary to point out that careful implementation of conditions like to $F(\mathbf{x}_i) > F(\mathbf{x}_k)$ might further improve the efficiency of the proposed algorithm, because of comparison operation is the longest operation after division, see tab.1.

## 4. Conclusion

The new efficient algorithm of $O(\lg N)$ complexity for clipping lines against convex window in $E^2$ has been developed. Edges of the given convex polygon can be arbitrarily oriented. It also proved the applicability of Computational Geometry results [CHA87a] even for small N. Similarly as the Rappaport's algorithm the proposed algorithm can be easily modified for polygon clipping. The suggested algorithm also proved the duality principle with the problem point-in-polygon, see [PRE85a], [NIE92a], [NIE92b]. It also proved applicability of principles of Computational Geometry results [CHA87a] even for small N. Similarly as Rappaport's algorithm the proposed algorithm can be modified for polygon clipping, where the clipped polygon might be non-convex. Superiority of the proposed algorithm over CB, ECB and Rappaport's algorithms was proved by theoretical estimations and experimental results.

All tests were implemented in Borland C++ on PC 486/33 MHz 256KB Cache. It is expected that for workstations the efficiency n will be higher than for PC 486 as the comparison operation is the longest operation used in the algorithm, see tab.2, and the timing ratio of operations on workstations is be better.

## 5. Acknowledgments

The author would like to express his thanks to students of Computer Graphics courses at the University of West Bohemia in Plzen and Charles's University in Prague who stimulated this work, especially to Mr.P.Bláha for careful test implementations and verification of the proposed algorithm, dr.A.Ferko and dr.F.Je ek for reading a manuscript, critical comments and suggestion they made and to anonymous referees who made critical and constructive recommendations that improved this paper very much.

## 6. References

[ABI90a] Abi-Ezzi,S.S., Wozny,M.J.: Factoring a Homogeneous Transformation for a More Efficient Graphics Pipeline, Computer Graphics Forum, Vol.9, No.3, pp.245-255, 1990.

[AKE91a] Akeley,K., Korobkin,C.P.: Efficient Graphics Processor for Clipping Polygons, US Patent No.5 051 737, 1991.

[AND89a] Andreev,R.D.: Algorithm for Clipping Arbitrary Polygons, Computer Graphics Forum, Vol.8, No.3, pp.183-192, 1989.

[AND91a] Andreev,R., Sofianska,E.: New Algorithm for 2- Dimensional Line Clipping, Computers & Graphics, Vol.15, No.4, pp. 519-526, 1991.

[ARO89a] Arokiasamy,A.: Homogeneous Coordinates and the Principle of Duality in Two Dimensional Clipping, Computers & Graphics, Vol.13, No.1, pp.99-100, 1989.

[BLI78a] Blinn,J.F.,Newell,M.E.: Clipping Using Homogeneous Coordinates, Computer Graphics (SIGGRAPH'78), Vol.12, pp.245-251, 1978.

[BLI91a] Blinn,J.F.: A Trip Down to Graphics Pipeline - Line Clipping, IEEE Computer Graphics and Applications, Vol.11, No.1, pp.98-105, 1991.

[BRE92a] Brewer,E.A., Barsky,B.A.: Clipping After Projection: An Improved Perspective Pipeline, submitted for publication.

[BUR88a] Burkert,A., Noll,S.: Fast Algorithm for Polygon Clipping with 3D Windows, Eurographics'88 Proceedings, pp.405-419, 1988.

[CHA87a] Chazelle,B., Dobkin,D.P.: Intersection of Convex Objects in Two and Three Dimensions, JACM, Vol.34, No.1., pp.1-27, 1987.
[CHA92a] Chazelle,B.: An Optimal Algorithm for Intersecting Three-Dimensional Convex Polyhedra, SIAM J. Computing, Vol.21., No.4., pp.671-696, 1992.
[CHE79a] Cheng,F., Yen,Y.: A Parallel Line Clipping Algorithm and Its Implementation, in CGI'89 Conference Proceedings, 1989.
[CYR79a] Cyrus,M.,Beck,J.: Generalized Two and Three Dimensional Clipping, Computers&Graphics, Vol.3, No.1, pp.23-28, 1978.
[DAY91a] Day,J.D.: A Comparisons of Line Clipping Algorithms, Queensland Univ. of Technology, School of Computing Science, Report 2-91, 1991.
[DAY91b] Day,J.D.: A New Two Dimensional Line Clipping Algorithms for Small Windows, Queensland Univ. of Technology, School of Computing Science, Report 3-91, 1991.
[DAY92a] Day,J.D.: A New Two Dimensional Line Clipping Algorithms for Small Windows, Computer Graphics Forum, Vol.11, No.4, pp.241-245, 1992.
[DOR90a] Dorr, M.: A New Approach to Parametric Line Clipping, Computers & Graphics, Vol.14. Nos.3/4, pp.449-464, 1990.
[DUV90a] Duvanenko,V.J., Robins,W.E., Gyurcsik,R.S.: Improving Line Segment Clipping, Dr.Dobb's Journal of Software Tools, Vol.15, No.7, pp.36,38,40,42,44-5,98,100, 1990.
[DUV93a] Duvanenko,V.J., Robins,W.E., Gyurcsik,R.S.: Simple and efficient 2D and 3D Span Clipping Algorithms, Computers & Graphics, Vol.17, No.1, pp.39-54, 1993
[FON92a] Fong,D.Y., Chu,J.: A String Pattern Recognition Approach to Polygon Clipping, Pattern Recognition, Vol.23, No.8., pp.879-892, 1992.
[FUN90a] Fung,K.Y., Nicholl,T.M., Tarjan,R.E., Van Wyk,C.J.: Simplified Linear Time Jordan Sorting and Polygon Clipping Information Processing Letters, Vol.35, pp.85-92, 1990.
[HER88a] Herman,I.,Reviczky,J.: Some Remarks on the Modelling Clip Problem, Computer Graphics Forum, Vol.7, No.4, pp.265-272, 1988.
[HUB90a] Hubl,J.,Herman,I.: Modelling Clip: Some More Results, Computer Graphics Forum, Vol.9, pp.101-107, 1990.
[HUB93a] Hubl,J.: A Note on 3D-Clip Optimisation, Computer Graphics Forum, Vol.12, No.2, pp.159-160, 1993.
[KAI90a] Kaijian,S., Edwards,J.A., Cooper,D.C.: An Efficient Line Clipping Algorithm, Computers & Graphics, Vol.14, No.2, pp.297-301,1990.
[KIL87a] Kilgour,A.C.: Unifying Vector and Polygon Algorithm for Scan Conversion and Clipping, TR CSC/87/R7, Univ. of Glasgow, May 1987.
[KRA89a] Kramer,G.: Notes on the Mathematics of PHIGS Output Pipeline, Computer Graphics Forum, Vol.8, No.3, pp.219-226, 1989.
[KRA92a] Krammer,G.: A Line Clipping Algorithm and Its Analysis, Computer Graphics Forum (EG'92 Conference Proceedings), Vol.11, No.3, pp.C253-266, 1992.
[LIA83a] Liang,Y.D.,Barsky,B.A.: An Analysis and Algorithms for Polygon Clipping, CACM, Vol.26, No.11, pp.868-876, 1983.
[LIA84a] Liang,Y.D.,Barsky,B.A.: A New Concept and Method for Line Clipping, ACM Transaction on Graphics, Vol.3, No.1, 1984, pp.1-22.
[LIA92a] Liang,Y., Barsky, B.A.: The Optimal Tree Algorithm for Line Clipping, Technical paper distributed at Eurographics'92 Conference, Cambridge, 1992.
[MAI92a] Maillot,P.G.: A New, Fast Method For 2D Polygon Clipping: Analysis and Software Implementation, ACM Transaction on Graphics, Vol.11, No.3, pp.276-290, 1992.
[MAR89a] Margalit,A.,Knott,G.: An Algorithm for Computing the Union, Intersection or Difference of Two Polygons, Computers & Graphics, Vol.13, pp.167-183, 1989.
[MAX93a] Max,N.: Polygon Clipping - Response, CACM, Vol.36, No.1., pp. 115, 1993.
[NIC87a] Nicholl,T.M.,Lee,D.T.,Nicholl,R.A.: An Efficient New Algorithm for 2D Line Clipping: Its Development and Analysis, ACM Computer Graphics, Vol.21, No.4, pp.253-262, 1987.

[NIC91a] Nicholl,R.A., Nicholl,T.M.: A Definition of Polygon Clipping, Report No.281, Computer Sci. Dept.,Univ.of West Ontario, 1991.
[NIE92a] Nielsen,H.P.: An Intersection Test Using Dual Figures, Technical Report GKTR-0892, Dept.of Graphical Communication, Technical Univ. of Denmark, Lyngby, October 1992.
[NIE92b] Nielsen,H.P.: Line Clipping Using Semi-homogeneous Coordinates, submitted for publication in CGF, 1992.
[OBA89a] O'Bara,R.M.,Abi-Ezzi,S.: An Analysis of Modeling Clip, in EG'89 Conference Proceedings, pp.367-380, 1989.
[PIN91a] Pinedo,D.: Window Clipping Methods in Graphics Accelerators, IEEE Computer Graphics and Applications, Vol.11, No.3, pp.75-84, 1991.
[PRE85a] Preparata,P.F., Shamos,M.I.: Computational Geometry, An Introduction, Springer Verlag, 1985.
[RAP91a] Rappaport,A.: An Efficient Algorithm for Line and Polygon Clipping, The Visual Computer, Vol.7, No.1, pp.19-28, 1991.
[SHA92a] Sharma,N.C., Manohar,S.: Line Clipping Revisited: Two Efficient Algorithms based on Simple Geometric Observations, Computers & Graphics, Vol.16, No.1, pp.51-54, 1992.
[SKA89a] Skala,V.: Algorithms for 2D Line Clipping, in CGI'89 Conference Proceedings, pp.121-128, 1989.
[SKA89b] Skala,V.: Algorithms for 2D Line Clipping, in EG'89 Conference Proceedings, pp.355-367, 1989.
[SKA93a] Skala,V.: Algorithm for Line Clipping in E2 for Convex Window (in Czech), Algorithms'93 Conference Proceedings, Bratislava, 1993.
[SKA93b] Skala,V.: An Efficient Algorithm for Line Clipping by Convex Polygon, Computers & Graphics, Vol.17,No.4, pp.417-421, 1993.
[SKA93c] Skala,V.: An Efficient Algorithm for Line Clipping by Convex Polygon, Preprint No.38, University of West Bohemia, Plzen, 1993.
[SLA92a] Slater,M., Barsky,A.B.: 2D Line and Polygon Clipping Based on Space Subdivision, accepted for publication in The Visual Computer, 1993.
[SOB87a] Sobkow,M.S., Pospisil,P., Yang,Y.-H.: A Fast Two-dimensional Line Clipping Algorithm via Line Encoding, Computers & Graphics, Vol.11, No.4, pp.459-467, 1987.
[SPR68a] Sproull,R.F.,Sutherland,I.E.: A Clipping Divider, in: Proc. APFIS FJCC, 1968.
[SUT74a] Sutherland,I.E., Hodgman,G.W.: Reentrant Polygon Clipping, CACM, Vol.17, No.1, pp.32-42, 1974.
[THE89a] Theoharis,T.,Page,I.: Two parallel Methods for Polygon Clipping, Computer Graphics Forum, Vol.8, pp.107-114, 1989.
[VAT93a] Vatti,B.: Polygon Clipping - Response, CACM, Vol.36, No.1., pp. 115, 1993.
[YIN92a] Ying,D.N.: A New Algorithm for Polygon Clipping and Boolean Operations, Univ. of Zhejiang internal report, Hangzhou, China, 1991.
[YON91a] Yong-Kui,L.: A New Algorithm for Line Clipping by Convex Polygon, paper submitted for publication, 1991.
[ZAC89a] Zachristen,M.: Yet Another Remark on the Modeling Clip Problem, Computer Graphics Forum, Vol.8, pp.237-238, 1989.

[EDE90a] Edelsbrunner,H.,Mucke,E.P.: Simulation of Simplicity: A Technique to Cope with Degeneration Cases in Geometric Algorithms, ACM Trans. on Graphics, Vol.9, No.1, pp. 66-104, 1990.

**New**

[SAV90a] Savka,M.V.: A Polygon Clipping Algorithm, Programming and Computer Software, Vol.16, No.3, pp.114-117, 1990.
[SHI90a] Shi,K.J., Edwards,J.A., Cooper,D.C.: An Efficient Line Clipping Algorithm, Computers & Graphics, Vol.14, No.2, pp.297-301, 1990.

[WES89a] Weston,D.E.: Correlation After Asymetrical Clipping, Journal of the Acoustical Society of America, Vol.85, No.4, pp.1607-1611, 1989.

[TAN84a] Tang,Z.H., Sun,J.G., Chen,Y.J.: A Method for Clipping Arbitrary polygon Rapidly, First International Conference on Computers and Applicatons, pp.457-464, Beijing, China, 1984.